\documentstyle[12pt,fleqn,espcrc1,epsfig]{article}
\newcommand{\beq}{\begin{equation}}
\newcommand{\eeq}{\end{equation}}
\newcommand\ga{\raisebox{-.5ex}{$\stackrel{>}{\sim}$}}

\newcommand{\bea}{\begin{eqnarray}}
\newcommand{\eea}{\end{eqnarray}}

\title{\bf RHIC: From dreams to beams in two decades}
\author{Gordon Baym \\
        Department of Physics, University of Illinois at
        Urbana-Champaign\\ Urbana, IL 61801, U.S.A.}

\begin{document}

\maketitle

\date{}

\begin{abstract}
This talk traces the history of RHIC over the last two decades, reviewing the
scientific motivations underlying its design, and the challenges and
opportunities the machine presents.
\end{abstract}

\section{THE VERY EARLY DAYS}

    The opening of RHIC culminates a long history of fascination of nuclear
and high energy physicists with discovering new physics by colliding heavy
nuclei at high energy.  As far back as the late 1960's the possibility of
accelerating uranium ions in the CERN ISR for this purpose was contemplated
\cite{bm}.  The subject received ``subtle stimulation" by the workshop on
``Bev/nucleon collisions of heavy ions" at Bear Mountain, New York, organized
by Arthur Kerman, Leon Lederman, Mal Ruderman, Joe Weneser and T.D.~Lee in the
fall of 1974 \cite{bm}.  In retrospect, the Bear Mountain meeting was a
turning point in bringing heavy ion physics to the forefront as a research
tool.  The driving question at the meeting was, as Lee emphasized, whether the
vacuum is a medium whose properties one could change; ``we should
investigate," he pointed out, ``$\ldots$ phenomena by distributing high energy
or high nucleon density over a relatively large volume."  If in this way one
could restore broken symmetries of the vacuum, then it might be possible to
create abnormal dense states of nuclear matter, as Lee and Gian-Carlo Wick
speculated \cite{leewick}.

    The physics discussions at Bear Mountain focussed on astrophysical
implications of unusual states of matter such as pion condensates and Lee-Wick
matter in neutron stars, high energy cosmic rays, stable abnormal nuclei, as
well as fanciful applications, e.g., by A. Turkevich to manufacture and custom
tailor superheavy materials, even to make a high temperature superconductor,
and by G. Vineyard to use abnormal nuclei as active components in a breeder
reactor.  The worry engendered by the latter's suggestion that seeds of
unusual states could set off a global catastrophe was calmed by the
observation that ``Lee-Wick theory indicates that $10^8$ or $10^9$ [abnormal
superdense nuclei] have already been produced on the moon, and that the moon
is still there, albeit with large holes" -- an approach still invoked to
support the safety of high energy nuclear collisions \cite{buzna}.  Schemes
for accelerating heavy ion beams were also addressed:  H. Grunder described
possibilities of injecting heavy beams from the SuperHILAC into the Bevalac (a
project in fact completed in 1984), G. Cocconi mentioned thoughts at CERN of
transferring ions up to $^{16}$O from the PS into the ISR and eventually into
the SPS \cite{isr}.  Most important for RHIC was K. Prelec and A. van
Steenbergen's proposal of constructing a booster ring to inject fully stripped
ions with A $\ge$ 200 into the AGS.

    One should remember that at the time of the Bear Mountain meeting, the
idea of quark matter as the ultimate state of nuclear matter at high energy
density had not taken hold.  The asymptotic freedom of QCD had only been shown
the previous year \cite{gross}.  Rather, in addition to Lee-Wick abnormal
matter, possible states under consideration included the Hagedorn hadronic
resonance gas \cite{hagedorn}, as elucidated by Frautschi and Lee, Leung, and
Wang \cite{frautschi}, and mean field hadronic models, most recently that of
Walecka \cite{walecka}.  Although the concept of quark matter was mentioned as
early as 1970 by Itoh in the context of neutron stars \cite{itoh}, and
described before asymptotic freedom by Carruthers in 1973 \cite{carruthers} --
as ``quarkium, a bizarre Fermi liquid" -- the decisive step was the paper of
Collins and Perry the year following Bear Mountain \cite{collins}.  Their
motivation was to understand the equation of state of matter, as needed to set
an upper limit on the maximum mass of a neutron star, a problem just discussed
by Rhoades and Ruffini \cite{ruffini} using the Hagedorn hadronic equation of
state.  Their crucial realization was that ultrahigh temperature as well as
ultrahigh baryon density corresponded to the asymptotic regime of QCD, rather
than a hadronic regime, and thus the ultimate state would be a weakly
interacting "quark soup."

    Several other early meetings were seminal in the eventual conception of
RHIC, including the ``first workshop" on ultrarelativistic nuclear collisions
at Berkeley in May 1979 \cite {qmlbl}, the 1980 GSI Workshop \cite{bock}, the
1980 joint Japan-U.S. seminar at Hakone \cite{hakone}, Helmut Satz's meeting
in Bielefeld, which was instrumental in bringing theorists together to think
about ultrarelativistic collisions and quark matter \cite{satz1}, and the
second conference in the Quark Matter series at Bielefeld in 1982
\cite{satz2}.  Plans for the fixed target heavy ion facilities at the AGS
\cite{ags} and at CERN \cite{sps} were well under way by early 1983.

    The critical event in establishing RHIC was the open ``town" meeting of
the U.S.  Nuclear Science Advisory Committee (NSAC) at Wells College in
Aurora, New York, from July 11-15, 1983.  The role of NSAC, which advises both
the NSF and DOE, is to coordinate nuclear science policy in the United States.
In the Spring of 1983 the immediate job of the Committee, which was chaired by
John Schiffer and of which I was a member, was to write a five year long-range
plan for nuclear physics, and in particular to recommend the next major
construction project to follow the just approved 4 GeV electron accelerator,
CEBAF, at the future Jefferson Lab.  Rather than dividing nuclear physics by
experimental facilities, the Committee's approach was to study the basic
science questions:  nuclear symmetries, quarks and QCD in nuclei, extreme
states of nuclear matter, nuclei and the universe, etc.  I found myself
chairing the subcommittee on extreme states of nuclear matter.  The other
members of the subcommittee were Arthur Kerman and Arthur Schwarzschild, who
were on NSAC, as well as Miklos Gyulassy, Tom Ludlam, Larry McLerran, Lee
Schroeder, Steve Vigdor, and Steve Koonin.

    The main issue to be decided at Aurora was whether the next facility would
be a hadron or heavy ion machine.  The hadron machine was proposed by Los
Alamos as the successor to the meson factory, LAMPF.  This machine, LAMPF II,
would inject protons from LAMPF into a 16-32 GeV synchrotron to generate a
K-meson beam as well as pion, muon, neutrino and $\bar {\rm p}$ beams.  The
most specific heavy ion project was the VENUS accelerator at Berkeley, a two
ring superconducting accelerator for both fixed target and colliding beam
experiments.  As chair of the subcommittee on nuclear matter under extreme
conditions, I was to make the scientific case for pursuing heavy ions.  My
intention, at a talk I would give in the middle of the meeting, was to
conclude with a statement that ``the highest priority for the field is an
ultra-relativistic heavy ion collider [of] E/A $\ga$ 30 GeV in the center of
mass, with A up to uranium."  But then a remarkable bit of news arrived Monday
evening, the first day of the meeting; the High Energy Physics Advisory Panel
(HEPAP), which advises DOE on high energy facilities, had just decided to
abandon the problematic Colliding Beam Accelerator (CBA), the 400 GeV on 400
GeV proton collider at Brookhaven -- whose construction was well under way --
in favor of building the then named Desertron, eventually the SSC.  Our
subcommittee realized immediately the remarkable opportunity this decision
opened to nuclear physics, and in my talk Wednesday morning, I argued the
proposal to build a colliding beam heavy ion accelerator in the CBA tunnel
\cite{bjorken}.  With the next day's favorable vote of the attendees at the
meeting (27 to 11 with one abstention), RHIC -- although not so named yet --
had entered the conceptual stage.

    As Schiffer summarized the deliberations to Jim Leiss, the Associate
Director at DOE for High Energy and Nuclear Physics, and Marcel Bardon, the
Director of the Physics Division at NSF \cite{schiffer},
\begin{quote}
    Our increasing understanding of the underlying structure of nuclei and
of the strong interaction between hadrons has developed into a new scientific
opportunity of fundamental importance -- the chance to find and to explore an
entirely new phase of nuclear matter.  In the interaction of very energetic
colliding beams of heavy atomic nuclei, extreme conditions of energy density
will occur, conditions which hitherto have prevailed only in the very early
instants of the creation of the universe.  We expect many qualitatively new
phenomena under these conditions; for example a spectacular transition to a
new phase of matter, a quark-gluon plasma, may occur.  Observation and study
of this new form of matter would clearly have a major impact, not only on
nuclear physics, but also on astrophysics, high-energy physics, the broader
community of science and on the world at large.  The facility necessary to
achieve this scientific breakthrough is now technically feasible and within
our grasp; it is an accelerator that can provide colliding beams of very heavy
nuclei and with energies of about 30 GeV per nucleon.  Its cost can be
estimated at this time only very roughly as about 150-200 million dollars.
{\em It is the opinion of this Committee that such a facility should be built
by the United States expeditiously, and we see it as the highest priority new
scientific opportunity within the purview of our science.}
\end{quote}

\section{THE BEGINNINGS OF RHIC}

    As an immediate followup to the Aurora meeting, Arthur Schwarzschild and
Tom Ludlam of BNL convened a task force which met from August 22-24, 1983 to
begin to set the parameters of the future heavy ion collider \cite{taskforce}.
The members of the Task Force from outside BNL included J. Bjorken, C. Gelbke,
H. Gutbrod, A. Kerman, C. Leeman, L. Madansky, A. Mueller, I. Otterlund, A.
Ruggiero, L. Schroeder, G. Young, W. Willis, and myself.  At this stage all
the civil engineering, including a $^4$He refrigeration system for
superconducting magnets, was in place for the now abandoned CBA; the
challenge was to ``stuff a collider" into the pre-existing tunnel.

    The first issue was the maximum energy of the collider.  The leading
consideration was to achieve a ``clean" central rapidity region, i.e., with
small net baryon density.  Experiments at the ISR indicated that the
projectile and target fragmentation regions in pp collisions were two units of
rapidity wide; nuclear effects were expected possibly to double this number
\cite{frag}.  Thus a lower bound on the energy would be 50 GeV/A per beam.
(The ``about 30 GeV per nucleon" in the Schiffer letter, above, was based on
this requirement.)  A compelling reason to go to 100 GeV/A was the possibility
of producing high energy jets, and studying their propagation through the
nuclear collision volume.  This process, which is being realized at RHIC
today, remains important as the closest one can come to carrying out deep
inelastic scattering to probe the matter in the collision volume.  Indications
from cosmic ray experiments at the time, particularly the Si on Ag JACEE
event, were that energy densities would be of order several GeV/fm$^3$, an
estimate that has been well substantiated by subsequent collisions at the SPS.
Such energy densities were felt then, as now, to be adequate to produce a
quark-gluon plasma.

    The Task Force stressed the importance of the beams having a large dynamic
range in energy and mass number, to allow systematic studies with increasing
mass number of the projectile nuclei, over a range of energies from the future
SPS program (equivalent to 10 GeV/A on 10 GeV/A) on upward.  It also
recognized the need to be able to run pp and pA collisions, as well as AA, to
be able to study the onset of new collective physics with increasing size of
the projectiles.  Designing the machine for pp collisions presented a delicate
political issue, since one did not want to appear to be resurrecting the CBA.
On the other hand, the capability of colliding protons has enabled the
development of polarized proton beams for the RHIC spin program (or
$\stackrel{\textstyle\longrightarrow}{\cal RHIC}$, as it could be known).

    Achieving a large luminosity was not a critical issue, since the cross
sections for central events are so large.  The Task Force set a minimum
luminosity of $10^{25}$ cm$^{-2}$sec$^{-1}$, with the possibility of
eventually upgrading to $10^{28}$ cm$^{-2}$sec$^{-1}$ to study rare events.
As the experimentalists have now experienced, RHIC's initial luminosity of ten
percent of its design luminosity $2\times 10^{26}$ cm$^{-2}$sec$^{-1}$ already
produces a rather healthy event rate.

    The report of the Task Force also sketched out the rudiments of a physics
program, utilizing at least three intersection regions, with at minimum two
large solid angle detectors and one small solid angle experiment.  The RHIC
experimental program took fuller shape at a number of workshops, including
that in Berkeley in September 1984 \cite{schroeder} and at BNL in April 1985
\cite{haustein}.

    The Third Quark Matter meeting at BNL in September 1983 played a
particular role in building community support for RHIC.  The meeting was
permeated, as Allan Bromley noted at the Round Table discussion of prospects
for future experiments, with ``a sense of enthusiasm, excitement, \ldots, a
feeling of adventure in the air."  In my Concluding Remarks at the meeting, I
laid out what seemed at the time like a reasonable timetable for construction
of RHIC, starting, after all needed reviews, in October 1987, with first beams
in October 1992.  [Hans Gutbrod, I recall, immediately stood up and asked
whether it really had to be that long!] The formal RHIC proposal, which was
issued in August 1984, sketched out an even more optimistic timeline, with a
project start in October 1985, and first colliding beam tests in July 1990.
Little did we imagine! \cite{hunter}

\begin{figure}
\begin{center}
\epsfig{file=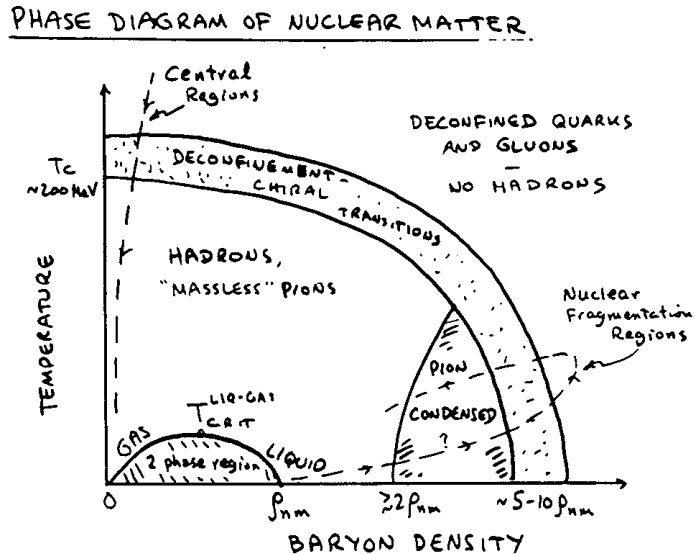, height=7.99cm}
\end{center}
\vspace{-.4cm}
    \caption{Phase diagram of nuclear matter in equilibrium, and how
it can be explored in ultrarelativistic heavy ion collisions, from the
1983 NSAC Long Range Plan \cite{nsaclrp}.}
\end{figure}


\section{THE SCIENTIFIC GOALS}

    The scientific base presented to NSAC in 1983 for carrying out
ultrarelativistic heavy ion collisions remains central in the goals of
RHIC today.  The basic questions asked were, ``What is the nature of nuclear
matter at energy densities comparable to those of the early universe?" and,
``What are the new phenomena and physics associated with the simultaneous
collision of hundreds of nucleons at relativistic energies?"  \cite{nsaclrp}
As the 1983 Long Range Plan put it, the most outstanding opportunity opened by
an ultrarelativistic heavy ion collider is ``the creation of extended regions
of nuclear matter at energy densities beyond those ever created in the
laboratory over volumes far exceeding those excited in elementary particle
experiments and surpassed only in the early universe."

    Now, as in 1983, nuclear matter at baryon densities well above nuclear
matter density, $\rho_{\rm nm}$, or at excitation energies corresponding
temperatures of hundreds of MeV is a {\it terra incognita}.  Knowledge of its
properties remains scant.  The proposed and probably naive equilibrium phase
diagram, reproduced from the 1983 Long Range Plan in Fig.~1, shows the
familiar low-temperature low-baryon density regime where the degrees of
freedom are hadronic, the high temperature or high baryon density regime where
matter is expected to be a quark-gluon plasma, and the uncertain transition
region between these two phases \cite{diagram}.  It also shows the expected
liquid-gas phase transition at low density, and a possible region of pion
condensed matter.  The regions of the diagram explored by the nuclear
fragmentation regions and the central region in ultrarelativitic collisions
are also shown.

    The heart of the program remains discovering the properties of nuclear
matter under extreme conditions.  Beyond simply mapping out its phase diagram
one would like to learn from ultrarelativistic heavy ion collisions
thermodynamic properties of high energy density nuclear matter, including its
entropy and equation of state, the nature of its excitations, e.g.,
quasiparticles and collective modes, how it transports energy-momentum,
baryons and other conserved quantities, how it emits particles, stops hadronic
and quark projectiles and otherwise dissipates energy.  These are tough
challenges, which will only be met with considerable theoretical modelling of
collisions.

    Much of the motivation for learning about dense matter has historically
come from astrophysics, and neutron stars in particular.  Such applications
were in the forefront at the Bear Mountain workshop, and at the NSAC Aurora
meeting, and were behind Collins and Perry's studies of the quark-gluon plasma
as the ultimate state of matter.  Understanding the properties of dense matter
remains crucial for determining the structure of neutron stars, their
mass-radius relation and thermal evolution, their upper mass limit and the
transition to black holes, as well as for answering the question of whether
there can exist a distinct family of quark stars.  They also enter in working
out how old stars undergo gravitational collapse and subsequent supernova
explosions.  New observations on compact x-ray sources, from the Rossi X-ray
Timing Explorer, the Chandra X-Ray Observatory, and other space telescopes,
are beginning to give information on strong field gravity as well as opening
the possibility of directly measuring neutron star masses in ``quasi-periodic
oscillation" objects (QPO's), thereby confronting theoretical expectations of
dense matter \cite{miller}.  However, while heavy ion collisions will provide
experimental information on hot dense matter-- potentially useful in studying
merging of binary neutron stars -- they will not directly measure properties
of cool matter in quasistatic neutron stars; learning about such matter will
require sufficient theoretical understanding of the hot regime to allow
extrapolation to sub-MeV temperatures.  Finally, the appeal of being able to
reproduce, in the central regions in ultrarelativistic collisions at RHIC,
conditions in the Big Bang at times from about one microsecond until the time
of nucleosynthesis, albeit under very dynamic rather than quasistatic
conditions, is irresistible.  Discovering how matter hadronized in the early
universe, whether via a sharp first or second order phase transition, or via a
crossover, and determining the associated entropy changes would be a
remarkable contribution of RHIC to cosmology.

    Ultrarelativistic heavy ion collisions continue to offer promise of
opening a new window on QCD, particularly on large distance scales not
reachable in few hadron collisions.  Over the last two decades the list of
questions that experiment can address, and possible answers, has become quite
refined.  A major advance has been the realization of the sensitive dependence
of the initial states formed in the collision volume to the partonic structure
of the incident nuclei.  How well, though, does the initial phase come to
local thermal equilibrium?  Can one measure and accurately predict the time
scales?  How can one extract the effective interactions between quark and
gluon degrees of freedom at distances of 5-10 fm?  How do the long-range
unscreened color magnetic interactions affect the structure of quasiparticles
in the plasma and their interactions?  Can one see evidence of a color
superconducting state?  What is the nature of the deconfinement transition?
How is fragmentation into hadrons affected by the presence of a dense cloud of
excitations, i.e., how does a quark-gluon plasma ``vulcanize" into hadronic
matter?  What is the role of chiral symmetry breaking in the transition; does
it lead to detectable disordered chiral condensates?  These are all issues
whose resolution will require considerable interplay between theory and
experiment.

    In addition, as the 1983 Long Range Plan noted, ultrarelativistic
collisions may produce a spectrum of unusual objects as the plasma expands and
hadronizes; the current list of hopefuls includes multiquark states, hadrons
with heavy quarks, extended droplets of large strangeness, and multi-baryon
states of unusual chiral topology.  Even pi-mu and other exotic atoms can be
formed in collisions.  But most importantly, we must remain prepared for
nature to surprise us in the way it reveals the physics of this unexplored
regime, as it did in first presenting neutron stars in the form of pulsars.

\section{THE AGS-SPS FIXED TARGET PROGRAMS}

    The AGS and SPS fixed target programs, which began experiments in 1986
have served very importantly as a warmup to RHIC; the carrying out and
analysis of the experiments, given the complexity of the final states, has
been a non-trivial feat of the nuclear and high-energy community.  Thanks to
the fixed target program, RHIC is not beginning in a vacuum; rather, the
experimentalists, as well as theorists, are battle-hardened from the AGS and
SPS.

    Beyond establishing the lay of the land in high energy collisions, the
fixed target experiments have produced tantalizing results.  The experiments
to date show clearly that many secondary interactions take place early on in
the collisions, producing behavior well beyond that seen in pp collisions.
Identification of directed and elliptic flow has shown that the dynamics in
the collision volume are collective \cite{flow}.  Inclusive measurements
together with careful analysis of two particle correlation (Hanbury
Brown--Twiss) data \cite{jacak} have given a detailed picture of the evolving
collision volume, and have shown that the experiments have produced matter at
unprecedentedly high energy density, an order of magnitude beyond that in
laboratory nuclei and certainly in the expected region for plasma formation.
The experiments to date have provided thermodynamic information on the early
stages of matter and freezeout conditions in the collisions \cite{pbm}.

    One of the first indications of new physics to emerge from the fixed
target program is the enhancement of strangeness compared with pp collisions,
first seen at the AGS by E802 in K$^+/\pi^+$ ratios \cite{E802}, and studied
in multistrange baryons at CERN, most recently, by NA49 and WA97
\cite{WA97}.  The second is the suppression of the J/$\psi$, as studied by
NA38 and then NA50 \cite{NA50}, which appears to defy explanations in terms of
nuclear absorption.  While it is very tempting to ascribe the suppression to
screening in a plasma, as proposed by Matsui and Satz \cite{ms}, we do not
fully understand how a nascent J/$\psi$ would be quenched in a hot strongly
interacting hadronic soup.  A further indication of unusual physics is the
excess of low mass dileptons observed by CERES and HELIOS/3 at CERN
\cite{ceres}, which points to a decrease of the rho mass in the hot stages of
the collision.

    Despite suggestive hints \cite{cernrpt}, the experiments have not yet
identified a quark-gluon plasma.  On the one hand, we do not understand the
strongly interacting hadronic state near the deconfinement transition well
enough to rule it out.  One cannot simply go from asymptotic cross sections to
deriving the properties of dense matter, and thus be able to assert with any
certainty that this is {\it not} the matter present.  Such an approach does
not work in condensed matter physics or in nuclear physics; one cannot, for
example, derive the properties of nuclei simply using nucleon-nucleon
scattering cross sections.  On the other hand, we do not yet understand the
quark-gluon plasma well enough to rule it {\it in}.  Pictures based on
perturbative QCD are neither trustworthy nor adequate at the energy densities
present.  Neither are lattice QCD calculations at a stage where they provide
accurate guidance, particularly since they have not yet dealt satisfactorily
with finite net baryon density, as is present in the fixed target experiments.

    It will be the role of future experiments at RHIC to characterize the
matter in the collisions.  To show that a quark-gluon plasma has been produced
will require providing evidence for color deconfinement, e.g., delineating the
effective degrees of freedom of the matter as those of quarks and gluons.
While creating and identifying a quark-gluon plasma is an exciting goal, in a
basic sense it is only one part of the larger question.  Matter created at
RHIC with effective interhadron separation much less than the diameter of
hadrons will, under any circumstances, be very different from standard nuclear
matter from the stockroom.  Whatever form such matter takes, it will be
interesting in its own right.  Its degrees of freedom will certainly not be
the familiar hadronic ones.  It may correspond to the simple theoretical
picture of a weakly interacting quark-gluon plasma; more likely it will be
intrinsically strongly interacting and, one should hope, much more complicated
and richer.  Discovery of the high energy phases of matter would only be the
beginning.  We should continue to bear in mind Kozi Nakai's question at the
1983 Quark Matter meeting in Brookhaven, "What is the next step after we find
it [the quark-gluon plasma]?"  \cite{nakai}

\section*{ACKNOWLEDGEMENTS}

    The beginning of the RHIC physics program gives us a good opportunity to
thank all the people who over the years have contributed so much to the
development of RHIC.  In addition to those mentioned in this brief history,
and other unsung principals from BNL and outside, we owe a special debt of
gratitude to the late Herman Feshbach, who understood and first set in motion,
particularly through the Nuclear Science Advisory Committee, the process in
the nuclear community that led to RHIC; to Nick Samios, who recognized from
the beginning the importance of RHIC and put the full resources of BNL behind
it; to Dave Hendrie whose support and guidance within DOE was crucial at all
times; and to Satoshi Ozaki for masterfully bringing RHIC to completion.

    Barbara Jacak's valuable comments on this paper are greatly appreciated.
This work has been supported in part by National Science Foundation Grant PHY
98-00978.



\begin{thebibliography}{99}
\bibitem{isr} Despite the appeal of the ISR to the heavy ion community,
the building of LEP would take priority in the eighties, with consequent
decommissioning of the ISR.
\end{thebibliography}
\end{document}